\documentstyle[12pt]{article}
\begin{document}

\begin{center}
{\LARGE{\bf Yang-Baxter equation and reflection\\
\vspace{1\baselineskip}
equations in integrable models}}
\end{center}

\vspace{2cm}

\begin{center}
{\large{P. P. Kulish \footnote{On leave of absence from the St.Petersburg's
Branch of Steklov Mathematical Institute of the Russian Academy of
Sciences, Fontanka, 27, St.Petersburg, 191011, Russia}}}
\end{center}

\vspace{2cm}

\begin{center}
{\small{\it Departamento de F\'{\i}sica Te\'orica and IFIC\\
Centro Mixto Universidad de Valencia - CSIC\\
46100 Burjassot (Valencia) Spain.}}
\end{center}

\begin{abstract}
The definitions of the main notions related to the quantum inverse
scattering methods are given. The Yang-Baxter equation and reflection
equations are derived as consistency conditions for the factorizable
scattering on the whole line and on the half-line using the
Zamolodchikov-Faddeev algebra. Due to the vertex-IRF model correspondence
the face model analogue of the ZF-algebra and the IRF reflection equation are
written down as well as the $Z_2$-graded and colored algebra forms of the YBE
and RE.
\end{abstract}
\newpage

\section{Introduction.}

Integrable and/or solvable models have been always quite important for
theoretical
physics. One of their attractive features is the direct possibility to go
beyond perturbation theory giving a solid background for theoretical
hypothesis and constructions. Another valuable characteristic is related to
the combination of different mathematical methods required to solve
some specific model
thus giving rise to mutual interrelations among formally separated
fields of mathematics. The start of the recent activity in the quantum
integrable models (see reviews \cite{1,2,3,4,5,6,7,8}
 and Refs. therein) was definitely
related with the development of the soliton theory, although the influence
of the preceding pioneering contributions is out of discussion (see
\cite{9,10,11,12} and Refs. therein).
This is a large field of research and even
recent monographs \cite{10,11,12,13,14,JM, 15,58}
do not overlap much in between.

There is variety of quantum integrable models and quite a few interrelations
among them. In these lectures we restrict ourselves to those models
the integrability of which is related more or less directly to the Yang-Baxter
equation and to reflection equations. More of that, we will discuss mostly
the algebraic properties of these equations, their solutions and corresponding
integrable models putting aside very elaborated analytical techniques and/or
problems (of great physical significance) such as the thermodynamic limits of
finite size systems, massless and/or conformal field theory limits, continuous
limits of the lattice models, correlation
functions and form factors, critical exponents etc. The main notions
related to
the Yang-Baxter equation (YBE) will be introduced: factorizable scattering on
the line, the Zamolodchikov-Faddeev algebra, the fusion procedure or the
bound state scattering, the Yang-Baxter or transition matrix algebra,
integrals of motion. The factorizable scattering on a half-line gives rise to
the reflection equation (RE) and a boundary operator. Most of the applications
of the YBE can be extended to the RE case with appropriate modifications.
Due to the vertex-face model correspondence of statistical mechanics we
introduce an
interaction round face
analogue of the Zamolodchikov-Faddeev algebra and the corresponding
reflection equation for the boundary weights. The super or $Z_{2}$-graded
analogue of these constructions with the generalizations to
the color algebras will also be given as well as some particular
integrable systems with finite degrees of freedom. The interrelations among
few forms of the Bethe Ansatz: coordinate, algebraic, analytical and functional
will be mentioned as well.

More particular applications of the Bethe Ansatz technique to specific
integrable models with direct physical meaning
( a high energy limit of the QCD, the Azbel - Hofstadter problem )
and of the reflection equation
formalism to the Chern-Simons theory and quantized moduli spaces of flat
connections can be found in other lectures of this School.

\newpage

\section{Yang-Baxter equation and
Zamolodchikov-Faddeev algebra}

Let us start the presentation which will be full of definitions and
technicalities from a specific
integrable model to have in mind an example. This is the
one-dimensional Bose gas consisting of
$n$ sort particles with the Dirac delta-function two particle potential.
The field operators $\psi_{a} (x), \psi_{b}^{\dagger} (y)$
satisfy the canonical commutation relations, $a, b= 1,2, \cdots , n$

\begin{equation}\label{2.1}
[\psi_{a} (x), \psi_{b}^{\dagger} (y)] = \delta_{ab} \delta (x-y) \quad ,
\quad [\psi_{a} (x), \psi_{l} (y)]=0.
\end{equation}

\noindent
The field Hamiltonian is

\begin{equation}\label{2.2}
H= \int dx (\partial_{x} \psi_{a}^{\dagger} (x) \partial_{x} \psi_{a} (x) + c
\psi_{a}^{\dagger} (x) \psi_{b}^{\dagger} (x) \psi_{b} (x) \psi_{a}(x))
\end{equation}

\noindent
The translational invariance (on the whole line or with the periodic boundary
conditions, circle) and the internal $ U (n)$ invariance are obvious. The
corresponding symmetry generators are

\begin{equation}\label{2.3}
P= -i \int dx \psi_{a}^{\dagger} (x) \partial_{x} \psi_{a} (x), \quad
 U_{ab}=
\int dx \psi_{c}^{\dagger} (x) (u_{ab})_{cd} \psi_{d} (x).
\end{equation}

\noindent
where $u_{ab}$ are $n^{2}$ generators of the unitary Lie algebra $u (n)$. One
of these operators is the particle number operator

\begin{equation}\label{2.4}
N= \int dx \psi_{a}^{\dagger} (x) \psi_{a} (x) \; .
\end{equation}

The Fock space of states is the direct sum of the number operator $N$
eigenspaces

\begin{equation}\label{2.5}
{\cal H}_{F} = \sum^{\infty}_{M=0} {\cal H}_{M}
\end{equation}

\noindent
The common eigenfunctions $\Psi_{M}$ of operators $N, P$, and $H$ (2)

\begin{equation}\label{2.6}
\Psi_{M} = \int d^{M} x \Psi (1, \cdots , M|\lambda_{1},\cdots , \lambda_{M})
\prod_{i=1}^{M} \psi_{a_{i}}^{\dagger} (x_{i}) |0\rangle
\end{equation}

\noindent
are constructed as appropriate linear combinations of the one particle
eigenstates (the plane waves). The numbers $j=1,2,\cdots M$ in the argument of
$\Psi$ (6) refer to both coordinate $x_{j}$ and isotopic $a_{j}$ indices.
The coefficients in the linear combination depend on the one particle
parameters
(momenta) $\lambda_{j}$ and on the isotopic indices $a_{j}$

\begin{equation}\label{2.7}
\Psi ( \{ j \} | \{ \lambda_{j} \})= \sum_{\sigma \in {\cal S}_{M}} A_{\sigma}
(\{a_{j} \}, \{\lambda_{n} \}) exp (i \sum^{M}_{m=1} \lambda_{\sigma m} x_{m})
\end{equation}

\noindent
where $\sigma $ are all elements of the permutation group ${\cal S}_M $.

Such a form of the Hamiltonian eigenfunctions is known as
the {\bf coordinate Bethe Ansatz}.
The conditions of the wave function continuity and its appropriate derivative
jumps on the hyperplanes $x_{j} = x_{j + 1}$ (the sewing conditions) define
the coefficients $A_{\sigma} (\{a_{j} \} | \{ \lambda_{k} \})$.
The coefficients $A_{\sigma}$ and $A_{\sigma '}$ with
$\sigma '= \sigma_{j} \sigma$ where $\sigma_{j}$ is the transposition of the
indices $j, j + 1$, are related by the two particle $S$-matrix [20]:

\begin{equation}\label{2.8}
A_{\sigma '} = S (\lambda_{j} - \lambda_{j + 1}) A_{\sigma}\; ,
\end{equation}

\noindent
where for the model chosen (2)

\begin{equation}\label{2.9}
S (\lambda ) = (\lambda + ic{\cal P}) /(\lambda -ic)\; ,
\end{equation}

\noindent
and $\cal P$ is the permutation operator in $C^n \otimes C^n$.
The periodicity condition for the system on finite interval $(0, L)$ results
in the {\bf Bethe equations} for the set of $M$ momenta $\lambda_j$

\begin{equation}\label{2.10}
exp(i\lambda_{j}L)= - \prod_k S_{jk} (\lambda_{j} - \lambda_{k}) \; ,
\end{equation}

\noindent
where in the ordered product $k=j+1, ..., M-1, M, 1, ..., j-1$. Hence the
meaning
of the RHS is the scattering matrix of the $j$-th particle on the
other $(M-1)$ particles.
This would be just a phase factor for the scalar particles $(S(\lambda )=
(\lambda+ic)/(\lambda-ic) )$, but for the $n$ component case one has to
diagonalize the complicated scattering matrix to arrive to a system of
scalar equations. All quantities in the RHS are particular values
at $\lambda = \lambda_{j}$ of the {\bf transfer matrix}  $(k=1, 2, ...,M)$

\begin{equation}\label{2.11}
t(\lambda ; \{\lambda_m\}) = tr_a T(\lambda ; \{\lambda_m\})
\equiv tr_a  \prod_k S_{ak} (\lambda - \lambda_{k})
\end{equation}

\noindent
of the inhomogeneous $GL(n)$-spin magnet of $M$ sites with the transition or
monodromy matrix $T(\lambda ; \{\lambda_m\})$. The trace in the expression for
the transfer matrix $t(\lambda ; \{\lambda_m\})$ is taken over
the {\bf auxiliary space} $V_a = C^n$, while $t(\lambda ; \{\lambda_m\})$
is an operator (matrix) in the
space $ \prod_{k=1}^M (C^n)_k $. The important property of the
commutativity of the transfer matrix for different values of the spectral
parameter

$$
[t(\lambda ; \{\lambda_m\}), t(\mu ; \{\lambda_m\})] = 0
$$

\noindent
follows easily from the fundamental commutation relation for the transition
matrix $T(\lambda ; \{\lambda_m\})$ (see (15)).

Hence the RHS of (10) $t(\lambda_j ; \{\lambda_m\})$ can be diagonalize
simultaneously.
As result the {\bf hierarchy of the Bethe Ansatze} appears
(or the nested Bethe Ansatz) and the complete parametrization of
$\Psi ( \{ j \} | \{ \lambda_{j} \})$ has $n$
sets of "quasimomenta" including ${\lambda_j}$ (see below).

The consistency condition of this system is the {\bf Yang-Baxter equation}
(YBE) for the $S$-matrix $S (\lambda )$:

$$
S_{jk} (\lambda_{j} - \lambda_{k}) S_{jl} (\lambda_{j} - \lambda_{l}) S_{kl}
(\lambda_{k} - \lambda_{l})=
$$

\begin{equation}\label{2.12}
= S_{kl} (\lambda_{k} - \lambda_{l}) S_{jl} (\lambda_{j} - \lambda_{l})
S_{jk} (\lambda_{j} - \lambda_{k}).
\end{equation}

\noindent
One uses very often in the general setting the notation $R$
( the {\bf $R$-matrix} ) for the YBE solution.

The complete scattering matrix $S ( \{ \lambda_{k} \})$ of the $M$ particle
is given by the
ratio of the coefficients of the incoming wave
$(\lambda_1 < ...< \lambda_M )$ and the outgoing wave

$$
exp(i \sum_j \lambda_j x_j), \quad exp(i \sum_j \lambda_{M-j+1} x_j) \; .
$$

\noindent
This ratio is factorized into
the ordered product of $M (M-1)/2$ two particle $S$-matrices (9)
( the {\bf factorizable scattering} on the line ).

It was proposed for an algebraic description of the factorizable
scattering in the general case
to introduce a set of (annihilation) operators $Z_{a} (\lambda )$ \cite{16}
satisfying the commutation relations ( the {\bf Zamolodchikov algebra} )

\begin{equation}\label{2.13}
Z_{a} (\lambda ) Z_{b} (\nu) = S_{ab, cd} (\lambda - \nu) Z_{d}
(\nu) Z_{c} (\lambda ),
\end{equation}

\noindent
where $S (\lambda - \nu)$ is an $n^{2} \times n^{2}$ matrix. Using the
associativity property of this algebra and changing the order of the product
$Z_{a_{1}} (\lambda_{1}) Z_{a_{2}} (\lambda_{2})
 Z_{a_3 }(\lambda_{3})$ in
two possible ways one arrives to the consistency condition (12).
Extending the Zamolidchikov algebra (13) by adding $n$ more (creation)
conjugated operators $Z_{a}^{\dagger} (\nu)$ one gets
the {\bf Zamolodchikov-Faddeev algebra} (ZF-algebra)

\begin{equation}\label{2.14}
Z_{a} (\lambda ) Z_{b}^{\dagger} (\nu) = \delta_{ab} \delta (\lambda - \nu) +
Z_{c}^{\dagger} (\nu) \hat{S}_{ac,bd} (\lambda - \nu) Z_{d} (\lambda ).
\end{equation}

It is useful to write down the ZF-algebra in a compact matrix form by
introducing the
$n$ component column $A(\lambda)= (Z_1(\lambda), ..., Z_n(\lambda))^t$ and
the $n$ component row
$A^{\dagger}(\nu)=(Z_1^{\dagger}(\nu), ..., Z_n^{\dagger}(\nu))$. Then the
defining
relations of the ZF-algebra are

$$
A (\lambda ) \otimes A (\nu)
\equiv A_{1} (\lambda ) A_{2} (\nu) = S_{12} (\lambda - \nu) A_{2}
(\nu) A_{1} (\lambda ),
$$

$$
A_{1}^{\dagger} (\lambda ) A_{2}^{\dagger} (\nu) =
A_{2}^{\dagger} (\nu) A_{1}^{\dagger} (\lambda )
S_{21}^{\dagger} (\nu - \lambda) ,
$$

$$
A_{1} (\lambda ) \otimes A_{1}^{\dagger} (\nu) =
I_1 \delta (\lambda - \nu)+
A_{2}^{\dagger} (\nu) \hat{S}_{12} (\nu - \lambda) A_{2} (\lambda ),
$$

\noindent
where subscripts refer to the corresponding isotopic spaces $C^n \otimes C^n
\equiv V_1V_2 $ and $S_{21}= {\cal P}S_{12}{\cal P}, \;
\hat{S}_{12}={\cal P}S_{12}$.
Due to the unitarity property of the $S$-matrix:
$S_{12} (\lambda - \nu)S_{21} (\nu - \lambda) = I_{12}$ and the YBE one
can construct the Fock space representation ${\cal H}_F$ of the ZF-algebra
using the generalizing symmetrizing operators which include $\hat{S}$ instead
of the permutation operators [37].

To solve the non-linear Schr\"odinger equation (NS) ( the Heisenberg equation
of motion for $\psi_a(x,t)$ with the Hamiltonian (2) )
in the framework of the quantum inverse scattering method (QISM) [1 - 7]
one has to find the corresponding {\bf auxiliary linear problem}

$$
\frac{d}{dx} T (\lambda, x) = L (\lambda, x) T (\lambda, x)\, .
$$

\noindent
For the model in question $L$ depends on the spectral parameter $\lambda$
and the original dynamical local variables (1)

$$
L (\lambda, x) = \lambda J + \sum_a (c \psi_a (x) e_{a, n+1} +
\psi_a^{\dagger} (x) e_{n + 1, a})
$$

\noindent
where $J$ is $(n + 1) \times (n + 1)$ diagonal matrix:$J=$ diag $(I_{n} , - 1)$
and $e_{ij}$ are the $(n + 1) \times (n + 1)$ basis matrices. Hence the
auxiliary linear space here is $C^{n+1}$.
This is the {\bf L-operator} of the QISM or the classical soliton
theory [13]. Solution to the operator valued matrix first order equation
with normal ordering with respect to the local fields (1) and
the vacuum $ |0\rangle$

$$
T (\lambda, x) = : exp ( \int^x L (\lambda, y) dy):
$$

\noindent
defines the {\bf transition or monodromy matrix} $T (\lambda, x)$. Its
entries are the new variables of the model (the quantum
scattering data (QSD)).

For the NS case it is natural to represent $T (\lambda, x)$ in the block
form

$$
T = \left(
\begin{array}{cc}
A & B\\
C & D \\
\end{array} \right)
$$

\noindent
where $A$ is an $n \times n$ matrix, $B$ and $C$ are $n$ component vectors,
$D$ is a scalar. The quadratic commutation relations of the new
variables are defined by the fundamental relation \cite{1, 2, 3}

\begin{equation}\label{2.15}
R (\lambda - \nu) T_1 (\lambda) T_2 (\nu) = T_2 (\nu) T_1 (\lambda)
R (\lambda - \nu)
\end{equation}

\noindent
where the standard QISM notations [1-7] are used $T_1 = T \otimes I, \;
T_2 = I \otimes T $ to embed matrices in $C^n$ into $ C^n \otimes C^n$.
{}From the structure of the $L$-operator one can conclude that the $n$ elements
of the row $C(\lambda)$ act on the vacuum as some creation operators. This is
the starting point to construct the eigenfunctions of the transfer matrix
$t(\mu) = tr T(\mu) = tr A + D $ algebraically:

$$
\Psi_{M} = \prod_{j = 1}^{M} C_{a_j} (\lambda_j) 0\rangle
$$

\noindent
in the framework of the {\bf algebraic Bethe Ansatz} \cite{1, 2, 3}.

The mentioned above commutativity of the transfer matrices $t(\mu) = tr T(\mu)$
follows from (15) taking the trace of $T_1T_2 = R_{12}^{-1}T_2T_1R_{12}$ over
both spaces $ C^n \otimes C^n$. The trace form of the integrals of motion
generating function $t(\mu)$ for continuous as NS (2) or chain models as (11)
gives rise to the periodic boundary conditions. For the $M$ site spin
Hamiltonian
$H = (\sum_{n = 1}^{M-1} h_{n,n+1}) + h_{M,1}$ the $M$-th site spin interacts
with the first one. The treatment of the non-periodic boundary conditions
in the algebraic framework of the QISM requires the RE [29, 32, 34, 31]
(for the coordinate Bethe Ansatz see [12, 28]).

Using block decomposition for the $R$-matrix one can rewrite the compact
form of the fundamental relations (15) in terms of the blocks $A, B, C$,
and $D$. Restricting to the case of finite number
of particle and after appropriate limit to the whole line one gets
the ZF-algebra realization in terms of the QSD [50]

$$
Z (\lambda) = (A)^{-1}(\lambda) B(\lambda),
\quad Z^{\dagger} (\nu) = C (\nu)D^{-1} (\nu) \, .
$$

The $R$-matrix of (15) satisfies the YBE

$$
R_{12}(\lambda - \mu) R_{13}(\lambda - \nu) R_{23}(\mu - \nu) =
R_{23}(\mu - \nu) R_{13}(\lambda - \nu) R_{12}(\lambda - \mu) \, .
$$

\noindent
For the NS the $R$-matrix has the same structure as (9)
(but it is $(n + 1)^2 \times (n + 1)^2$ matrix).

In the general situation the $R$-matrix depends on the spectral parameter $u$
and some
other parameters $R (u; \eta, ...)$. Although there is no complete mathematical
theory of the Yang-Baxter equation, variety of solutions are known
as well as different fields of their applications. In particular, many
solutions
are related to the simple Lie (super) algebras. They are classified by the Lie
algebra, its irreducible representations $\Lambda_j$ and the spectral parameter
dependence: rational, trigonometric and elliptic ones [3, 9, 11, 48]. The
$sl(2)$
spin $1/2$ $R$-matrix related to the $XXX$-magnet
is used in few lectures of these Proceedings. The recent development relates
the spectral parameter dependent $R$-matrices with
the affine Lie (super-)algebras \cite{JM}. There are
also solutions to the YBE with the spectral parameter on the algebraic curves
of higher genus ( the Potts models ). The $R$-matrices acting in infinite
dimensional spaces can be found in papers [58].

Some particular properties of the YBE solution $R (u; \eta, ...)$,
which are important for different applications ( but not always valid for
a given solution ) are:
regularity

$$
R (0) = \phi (\eta) {\cal P}\; ,
$$

\noindent
$P$-symmetry

$$
{\cal P} R_{12} (u) {\cal P} \equiv R_{21} (u) = R_{12} (u) \; ,
$$

\noindent
$T$-symmetry

$$
R^{T}_{12} (u) = R_{12} (u) \, ,
$$

\noindent
unitarity

$$
R_{12} (u) R_{21} (- u) = \rho (u) I \; ,
$$

\noindent
crossing symmetry

$$
R^{t_1}_{12} (u) R_{12}^{t_1} ( - u - \eta) = \xi (u) I \, ,
$$

\noindent
quasiclassical property

$$
R (u; \eta) = I + \eta r (u) + {\cal O} (\eta^2)\, ,
$$

\noindent
where $r (u)$ is the classical r-matrix \cite{3, 12} and
$\phi (u), \; \rho (u), \; \xi (u) $ are some functions related to the
$R$-matrix
normalization. Many $R$-matrices have only $PT$-symmetry:
$R^{T}_{12} (u) = R_{21} (u)$.
The regularity is used to extract from $t(u)$ of the lattice
models the integrals of motion which are local in terms of initial
spin variables when the $R$-matrix itself is the $L$-operator.
The quasiclassical property gives rise to the direct connection of the quantum
model to the corresponding classical one \cite{3,12}.

It is easy to see that the product of the $R$-matrices $R_{13}(u)R_{23}(v)$ and
$R_{1'3}(w)$ are intertwined (like (15)) by the
matrix $R_{11'}(u - w)R_{21'}(v - w)$.
If the original $R$-matrix $R_{12}(x)$ is degenerated into a projector at
$x = \eta $ then one can project the above product to the corresponding
subspace for the fixed difference of the spectral parameters $u - v = \eta $.
For the Yang solution (9) at $\lambda = \pm ic$ one has
the ( anti )symmetrizer $P_+ ,\; P_-$ .
This is  the {\bf fusion procedure} \cite{3} to get new $R$-matrices from the
known ones. It has a direct physical interpretation as construction of the
bound
state $S$-matrix \cite{16,17}. Using the fusion procedure the $L$-operators in
the higher dimensional irreducible representations can be obtained giving
rise to the ZF operators for the bound states and to the integrable lattice
models of higher spins such as the spin $s$ $XXZ$-model.  The connection
of the $R$-matrices and integrable models with the simple Lie (super-)algebras
is
reflected in the structure of the Bethe equations: they include $r$ sets
of "quasimomenta" , where $r$ is the Lie algebra rank, and the Cartan matrix
[37, 47].

Omitting the spectral parameter dependence in the YBE one gets still very
interesting equation solutions of which ( the {\bf constant $R$-matrices} )
can be considered as structure constants of the quantum groups and the quantum
algebras. Then the quadratic relations (15) ( without the spectral parameter )
are the defining relations on the $n^2$ generators $T_{ab}$
of the corresponding quantum group.
The constant $R$-matrices $\hat{R}_{ij}$ have also direct relation to the
braid group (BG), for one of the defining relations of the BG generators
$\sigma_{i} \, \sigma_{i+1} \, \sigma_{i} =
\sigma_{i+1} \, \sigma_{i} \, \sigma_{i+1}$
coincides with the YBE for $\hat{R}_{ij} = \cal P_{ij} R_{ij}$:

$$
\hat{R}_{i-1i}\; \hat{R}_{ii+1}\; \hat{R}_{i-1i} =
\hat{R}_{ii+1}\; \hat{R}_{i-1i}\; \hat{R}_{ii+1} \;.
$$

\section{Reflection equations and
their covariance}

Let us consider factorizable scattering of particles with internal degrees of
freedom on a half-line [22]. Then even one-particle process in nontrivial
(a reflection from the wall) and it is described by an $n \times n$ matrix
$K (u)$ the {\bf reflection matrix} or the boundary $S$-matrix.

For an algebraic description it is useful to add to the ZF-algebra a
formal boundary operator $B$ with relation [24]

$$
Z_a (u) B = K_{ab} (u) Z_b (-u) B \, .
$$

\noindent
Then the two particle factorizability gives rise
additionally to the YBE for the two body $S$-matrix (12)
to the {\bf reflection equation} (RE)

\begin{equation}\label{3.1}
\begin{array}{cc}
S_{12} (u - v) K_1 (u) S_{21} (u + v) K_2 (v) = \\
& K_2 (v) S_{12} (u + v) K_1 (u) S_{21} (u - v) \, .
\end{array}
\end{equation}

\noindent
There are natural properties of the reflection matrix $K (u)$ as in the
preceding
Sec. of $R(u; \eta,...)$ :
$K(0) = I$ (regularity); $ K (u) K (-u) = I $ (unitarity); $T$-symmetry
$K^T(u)=K(u)$; the crossing symmetry is more elaborated and it
involves the $S$-matrix itself [24]. The constructions of the quantum group
invariant spin systems [30] uses the RE [31, 32, 42, 54].
Recent field theoretical applications of the RE can be found in [23-27, 36].

It is interesting to point out that as a consequence of the RE  (\ref{3.1}) the
quadratic combination

$$
\varphi (u) = \sum_{a,b} Z_a(-u) K_{ab} Z_b(u)
$$

\noindent
is a "local" field : $[\varphi (u),\; \varphi (w)] = 0$ \cite{19, 64}.
This property  was used to construct the metric tensor field for the quantum
Liouville theory \cite{64} with a slightly different RE and to propose a form
of the boundary operator $B = exp (\int_{0}^{\infty} \varphi (u) du)$ in the
dual
Hamiltonian picture of the boundary conformal field theory \cite{19}.

Although for many $R$-matrices solutions to the reflection equations were
found (cf [22, 25, 32, 33] ) there is no direct relation of them to the Lie
algebra theory. In particular, many of them do not depend on the quasiclassical
parameter $\eta $ (in the fundamental representation for the $R$-matrix).

Let us give two examples of the RE solutions.
Due to the $GL(n)$ symmetry of the Yang solution (9):
$[R,M \otimes M], M \in GL(n)$ the corresponding $K$-matrix can be transformed
$K \rightarrow K'=MKM^{-1}$ with arbitrary $M$ and the solution to the RE is

$$
K(u) = \xi I +uC, \quad C^2=I.
$$

In the elliptic case (the eight-vertex R-matrix [11]) the solution $K (u)$ is
[22, 25, 33]

$$
K (u) = \left(
\begin{array}{cc}
x (u) & y (u) \\
z (u) & w (u)
\end{array} \right) \, ,
$$

$$
\begin{array}{cl}
x (u) = & sn (\xi + u)\\[2ex]
y (n) = & \mu \, sn 2u (\lambda (1-k sn^2 u) + 1 + k sn^2 u)
(1- k^2 sn^2 \xi sn^2 u)^{-1}\,,\\[2ex]
w (u) = & sn (\xi - u )\, , \\[2ex]
z (n) = & \mu \, sn 2u (\lambda - 1 - (\lambda + 1) k sn^2 u ) (1 - k^2 sn^2
\xi sn^2 u)^{-1}\, ,
\end{array}
$$

\noindent
where $sn u \equiv sn (u ; k)$ is the Jacobi elliptic function of
modulus $ 0 < k < 1$, $\xi , \lambda , \mu$ are parameters.

The RE (16) has an important {\bf covariance property}: if $T(u)$ and $K(u)$
satisfy
the relations (15), (16) then $K'(u) = T(u) K(u) T(-u)^{-1}$ is also the RE
solution provided that the entries of $K(u)$ and $T(u)$ commute
$[K_{ab}(u),T_{cd}(u)]=0$. The proof follows easily by the substitution of
$K'(u)$ into the RE and using few times different forms of the fundamental
relation (15) e.g.

$$
T_2^{-1} (-v)R_{12} (u+v)T_1 (u) = T_1 (u) R_{12} (u+v) T_2^{-1} (-v).
$$

\noindent
This property gives rise to the {\bf Sklyanin monodromy matrix} \cite{25, 31}

\begin{equation}\label{3.2}
{\cal T}(u) = T(u) K(u) T(-u)^{-1} .
\end{equation}

\noindent
If the matrix $T(u)$ is constructed as an ordered product of $N$ independent
$L$-operators then $\cal T (u)$ can be interpreted as the monodromy matrix
of $N$ site lattice model with a boundary condition described
by the matrix $K(u)$ or a boundary interaction described
by the operator valued entries of the matrix $K(u)$. To extract the
corresponding
Hamiltonian and other integrals of motion the transfer matrix is constructed
using a special trace

\begin{equation}\label{3.3}
\tau (u) = tr K_{+}(u) T(u) K(u) T(-u)^{-1} =tr K_{+}(u) {\cal T}(u) \;.
\end{equation}

\noindent
An extra $K$-matrix $K_{+}(u)$ is any solution of a "conjugated" RE
[29, 32, 34, 51] defined in such way to guarantee the commutativity
$[\tau (u), \; \tau (w)] = 0$. In the regular case

$$
R_{an}(0) \sim {\cal P}_{an}\;, \quad K (0) \sim I\;, \quad
h_{n,n+1} =( \partial/ \partial u \hat{R}_{n,n+1} ) (0)\; ,
$$

\noindent
the Hamiltonian is [29]

\begin{equation}\label{3.4}
H = \sum_{n = 1}^{M-1} h_{n,n+1} + K_{1}'(0) +
(tr_{0} K_{+}(0)h_{M,0}) /tr K_{+}(0) \;.
\end{equation}

The structure of the {\bf fusion procedure for the reflection equation} is
similar to the $R$-matrix case, but the projected combination
( $R_{12} (\eta) \sim$ projector )

$$
K_1 (u) R_{21} (2u - \eta) K_2 (u - \eta)
$$

\noindent
includes the additional $R$-matrix in between the $K$-matrices [32, 51].

More general RE with four different $R$-matrices ( with or without the spectral
parameter )

$$
{R}_{12}^{(1)}\; K_{1} \; {R}_{12}^{(2)}\; K_{2}\;
= K_{2}\;{R}_{12}^{(3)}\;K_{1}\;{R}_{12}^{(4)}\; .
$$

\noindent
which are related among themselves by some consistency conditions
similar to (12) and (15), can be found in different papers
these days (see [55, 51] and Refs therein ). In particular, this kind of RE
was used for the quantum group invariant spin models with topological
interaction [42, 43]. Some generalizations of the RE are related to the
Coxeter groups [44].

The constant RE is also of interest as it was in the YBE case.
The corresponding RE-algebras with the $K$-matrix entries considered as
the generators, are related to the quantum group homogeneous spaces
while the $c$-number solutions $K$ can be considered as representations of
a special generator of the BG$^{(1)}$ in
the solid handlebody \cite{63} with the defining relation

$$
\hat{R}_{12}\; K_{1} \; \hat{R}_{12}\; K_{1}\;
= K_{1}\; \hat{R}_{12}\;K_{1}\; \hat{R}_{12}\; .
$$

In solvable models of statistical mechanics the $R$-matrix defines the
Boltzmann
weights
$R_{\alpha \beta;\gamma \delta} (u; \eta, ...)$ of the vertex models, where
$\alpha, \beta, \gamma$ and $\delta$ are spin variables on the four edges round
a vertex.  The models on the dual lattice
are known as the interaction round face models (IRF).
The corresponding Boltzmann weights $w(a, b, c, d|u)$
satisfy the {\bf star-triangular equation (relation)} (STR)
or Baxter relation \cite{10}

$$
\sum_{g} w (a, b, g, f | u) w (f, g,  d, e | u + v) w (g, b, c, d |v)=
$$

$$
= \sum_{g} w (f, a, g, e | v) w (a, b, c, g |  u + v) w (g, c, d, e | u)
$$

\noindent
The four sites surrounding a face are ordered anticklockwise from the
south-west corner and $u$ is a complex (spectral) parameter.

Let us introduce a face analogue of the ZF-algebra: Its generators are
parametrized by two indices $Z_{ab} (u)$ and satisfy the relations
(see e.g. \cite{81})

$$
Z_{ab} (u) Z_{bc} (v) = \sum_d w (a, b, c, d | u - v) Z_{ad} (v) Z_{dc} (u)
$$

\noindent
with the only one summation over $d$ in the RHS. As in the previous case the
scattering of a one particle $Z_{da} (u)$ on $N$ others $Z_{b_i c_i} (v_{i})$
will give a transition matrix

$$
T (u, v_{1}, \cdots , v_{N}) = \prod^{N}_{j=1} w (b_{j}, b_{j+1}, c_{j+1}\, ,
c_{j} | u - v_{j}).
$$

\noindent
by analogy with the YBE and ZF algebra constructions one can formulate the
following problems: 1) what are the conjugated (annihilation) operators and
their commutation relations with $Z_{ab}(u)$? 2) how to translate the fusion
procedure to the  STR  case? 3) what is an IRF analogue to the  RE ?

Let us start from the last question and introduce a formal operator $B$ of the
boundary which satisfies relations with $Z_{ab}(u)$:

$$
Z_{ab} (u) B = Q (a, s, b; u) Z_{as} (-u) B\, ,
$$

\noindent
where $Q (a, s, b; n)$ is a face analogue of the reflection matrix $K$. To
derive an  STR  analogue of the reflection equation the two
ways of transforming the product of three operators may be considered

$$
Z_{ab} (u) Z_{bc} (v) B \longrightarrow Z_{ak} (- u) Z_{kd} (-v) B
$$

\noindent
The resulting relation for the "scattering" matrix $w (\cdots \; )$ and the
reflection matrix $Q (\cdots \; )$ is \footnote{M.D. Gould informed me that the
IRF analogue of the RE was considered by P.A. Pearce.}

$$
w (a, b, c, g|u-v) Q (g, c, f|u) w (a, g, f, k| v + u) Q (k, d, f | v) =
$$

$$
= Q (b, m, c|v) w (a, b, m, s|u + v) Q (s, d, m|u) w(a, s, d, k|u-v)\;,
$$

\noindent
with summation over $g, f$ in the  LHS  and over $m,s$ in the  RHS .

To relate vertex models with the IRF models the Baxter intertwining
vectors come to play \cite{10}

$$
\sum_{\beta, \nu} R_{\beta \nu ; \lambda \nu} \phi_{ll'} (\beta)
z_{m' l'} (\nu) = \sum_{m}  w (m, m', l, l') z_{ml}
(\lambda) \phi_{m m'} (\alpha) \; .
$$

\noindent
These vectors remind a combination of the operators of the ZF-algebras
of the YBE and STR types. These vectors can be used to relate the reflection
matrices of the vertex and IRF models:

$$
\bar{\phi}_{ab} (\beta) K_{\beta \nu} (u) \phi_{bs} (\nu) = Q (a, s, b; u)\; ,
$$

\noindent
where the spectral parameter dependence of the vectors is omitted.

%\newpage

\section{Integrable models with anticommuting variables (fermions)}

One of the first integrable models of the quantum field theory-massless
Thirring model  had attracted a lot of attention of theoretical physicists
during decades

$$
H= \int dx ( \bar{\psi} \gamma \partial \psi + g (\bar{\psi} \psi )^{2})\; .
$$

\noindent
It contains Fermi fields $\psi_{\alpha} (x, t)\; , \; \psi_{\beta}^{+}
(x,t)$
satisfying anticommutation relations

$$
\{ \psi_{\alpha} (x, t), \psi_{\beta}^{+} (y, t)\} = \delta_{\alpha \beta}
\delta (x-y) \; .
$$

\noindent
The addition of the mass term $m \bar{\psi} \psi$ to the Hamiltonian gives rise
to the very rich dynamical properties of the massive Thirring model. In
particular,
it is dual to the famous sine-Gordon model with the change of the strong
coupling regime to the weak coupling one.

The famous $\delta$-potential Bose gas model [20] ( another name is quantum
nonlinear Schr\"odinger equation ) (2) was also
generalized to the multicomponent fermionic case in the
framework of the coordinate Bethe Ansatz [45] (see below).

Taking into account the "superization" procedure of the theoretical physics in
the seventies, let us follow this pattern and extend most of the preceding
equations and constructions to the case of the anticommuting variables.
Mathematically it is related with $Z_{2}$-graded vector space, $Z_{2}$-graded
algebras or as in the theoretical physics text: super-spaces, super-algebras,
super-analysis...

The $Z_{2}$-graded Zamolodchikov algebra is generated by the $(m + n)$
operators
$Z_{i} (u)$ satisfying

\begin{equation}\label{4.1}
Z_{i} (u) Z_{j} (v) = (-)^{p(i) p(j)} R_{ij;kl} \; (u, v) Z_{l} (v)
Z_{k} (u)\; .
\end{equation}

\noindent
where $i, j= 1,2, \cdots , m + n, p (i)$ is a parity function: $p (i)= 0\;
(i = 1, \cdots , m)$ and $ p (i)=1 \; (i=m + 1, \cdots , m + n)$,
hence one has $m$
boson and $n$ fermion operators. The coefficients in (\ref{4.1})
are $c$-numbers and the $R$-matrix is even:

\begin{equation}\label{4.2}
p (R_{ij;kl})= p (i) + p(j) + p(k) + p(l)=0\;.
\end{equation}

\noindent
Due to the $Z_{2}$ valuedness of the function $p(i)$,
and the $R$-matrix structure (\ref{4.2})
one can change sign factor in (\ref{4.1}) to
$(-)^{p(k)p(l)}$. The super-associativity requirement for
$Z_{a} (u) Z_{b} (v) Z_{c} (w)$ results in the graded Yang-Baxter equation
(grYBE)

\begin{equation}\label{4.3}
R_{ab;a'b'} \; (u,v) R_{a'c;j c'} \; (u,w)
R_{b'c'; hl} \; (v,w) \;(-)^{p(b')(p(j) + p(a'))}=
\end{equation}

$$
= R_{bc;b'c'} \; (v,w) R_{ac'; a'l} \; (u,w) R_{a'b';jk} \; (u,v)
(-)^{p(b')(p(a) + p(a'))} \;,
$$

\noindent
where the condition (\ref{4.2}) was used to reduce the factors in the LHS and
RHS:

$$
(-)^{p(a)p(l)+ p(a') p(c)+ p(b')p(c')}; \quad
(-)^{p(b)p(c)+p(a)p(c')+p(a')p(b')}\quad .
$$

\noindent
Recalling these sign factors one can write the grYBE in the same matrix
form as previously but now in the $Z_{2}$-graded tensor product of three
super-spaces $C^{m|n}$

\begin{equation}\label{4.4}
R_{12}\; (u,v) R_{13}\; (u,w) R_{23}\; (v,w) =
R_{23}\; (v,w) R_{13}\; (u,w) R_{12}\; (u,v) \: .
\end{equation}

\noindent
Considering the third space as an unspecified quantum space (or substituting
instead of the $(m + n) \times (m + n)$ blocks in $R_{13}$ and $R_{23}$
formal entries) one arrives to the graded FRT-relation

\begin{equation}\label{4.5}
R_{12}\; (u,v) T_{1}\; (u) T_{2}\; (v) =
T_{2}\; (v)  T_{1}\; (u) R_{12}\; (u,v)\quad .
\end{equation}

\noindent
The latter relation has the same sign factors in the component form as
(\ref{4.3}). These factors can be interpreted as result of the $Z_{2}$-graded
tensor product of two even matrices $F$ and $G$ ($p(F_{ac})= p(a)+p(c) $)

\begin{equation}\label{4.6}
(F \otimes G)_{ab;cd}= (-)^{p(b)(p(a)+p(c))} F_{ac} G_{bd}\; .
\end{equation}

\noindent
Hence, one gets additional sign factors for $T_{1}$ and the factor free $T_{2}$
(to consider more tensor factors the choice of $T_{1}$ factor free is
more convenient )

\begin{equation}\label{4.7}
(T_{1})_{ab;cd} = (T \otimes I)_{ab;cd} = (-)^{p(b)(p(a)+p(c))} T_{ac}
\delta_{bd}\; ,
\end{equation}

$$
(T_{2})_{ab;cd} = (I \otimes T)_{ab;cd}= \delta_{ac} T_{bd}
$$

\noindent
An additional sign factor appears also in the matrix of the permutation (flip)
operator

$$
{\cal P} (v \otimes w) = (-)^{p(v) p(w)} w \otimes v, \quad v, w \in C^{m|n}
$$

\begin{equation}\label{4.8}
({\cal P})_{ab;cd} = (-)^{p(a)p(b)} \delta_{ad} \delta_{bc}\; .
\end{equation}

\noindent
To arrive later to the graded reflection equation it is convenient to write
(\ref{4.7}) in a matrix form with the usual tensor product

\begin{equation}
(T_{1})_{gr} = \Gamma (T \otimes I) \Gamma ,
\end{equation}

\noindent
where $(\Gamma)_{ab;cd} = (-)^{p(b)p(a)} \delta_{ac} \delta_{bd}$, hence
$ \Gamma^2 = I $ (which is not the case of the color algebras). The
$Z_2$-graded
algebraic structures were analyzed also in [35] and many $R$-matrices were
found [48].

As in the first lecture the simplest integrable models correspond to the
rational solutions of the grYBE. The $GL(m|n)$ symmetric $R$-matrix is
( the graded Yang solution)

\begin{equation}\label{4.9}
R (u -v)=(u -v) + \eta {\cal P}.
\end{equation}

The corresponding models ($L$-operators) are the $GL (m|n)$ isotropic graded
magnets and the super-matrix nonlinear Schr\"odinger equation \cite{35, 65}

\begin{equation}\label{4.10}
L_{gm} (u)= u + \eta \sum_{i,j} e_{ij} \otimes s_{ji} (-)^{p(j)},
\end{equation}

\begin{equation}\label{4.11}
L_{NS} (x,u) = u J + \eta \sum (e_{ab} \psi_{ba} (x) \pm e_{ba}
\psi^{\dagger}_{ab} (x)),
\end{equation}

\noindent
where $s_{ij}$ are the generators of the super-algebra $gl (m|n),\; i,j = 1,2,
\cdots , m + n; \psi (x), \psi_{ab}^{\dagger} (x)$ are bose or Fermi fields
according to $p(a) + p(b)=0$ or 1 and $1 \leq b \leq N \,, \,
N + 1 \leq \, a \leq m + n$,  $J$ is
the block diagonal matrix $(I_{N} - I_{(m + n - N)})$.

Among integrable models related to the $Z_{2}$-graded case one can mention \\
$osp (1|2)$ - non-linear Schr\"odinger equation \cite{35}, which is described
by
the rational limit of the $osp (1|2) \; R$-matrix and the supersymmetric
sine-Gordon model with the trigonometric $sl (2|1) \; R$-matrix [3]. Although
the structure of the $osp (1|2) \; R$-matrix is similar to the one of the
$sl(2)$ spin 1 $R$-matrix the solution of the corresponding $osp (1|2)$-magnet
( or NS ) is obtained \cite{35} by the {\bf analytic Bethe Ansatz}. Due to the
arguments of analyticity, crossing symmetry and the bare vacuum the form of the
transfer matrix $t(u)$ eigenvalue is

\begin{equation}
\begin{array}{l}
\Lambda (u, \{v_j\}) = (u(u-d))^N \displaystyle{\prod_{k=1}^{M}}
S_1(u-v_k-1/2)S_{-1}(u-v_k-1) - \\
-((u-1)(u-d))^N \displaystyle{\prod_{k=1}^{M}} S_1(u-v_k) -
(u(u-d + 1))^N \displaystyle{\prod_{k=1}^{M}} S_{-1}(u-v_k-d) \, ,
\end{array}
\end{equation}

\noindent
where $S_l (u)= (u+l/2)/(u-l/2)$ and $d = 3/2$ . The regularity condition of
$\Lambda (u, \{v_j\})$ in $u$ ( the Manakov's principle ):
$Res \Lambda (u, \{v_j\}) = 0\,$ at $u = v_k$ gives the Bethe equations for the
set
of the quasimomenta $\{v_j\}$

$$
(S_1(v_k))^N = \prod_{j=1}^M S_2(v_k-v_j)S_{-1}(v_k-v_j) .
$$

\noindent
Hence the $M$ super-particle eigenstates are parametrized by one set of the
quasimomenta according to the Lie super-algebra rank one.

Another way to define the form of the reflection equation different from the
factorizability requirement is related with the covariance properties of the
$K$-matrix: if $K$ satisfies the RE then the same is true for the
transformed $K' = T K T^{-1}$. It follows from (\ref{4.5}), (28)

\begin{equation}
R_{12}\; (u,v) \Gamma K_{1} \; (u)\Gamma R_{21}\; (v,- u) K_{2}\; (v) =
\end{equation}

$$
= K_{2}\; (v)R_{12}\; (u,- v) \Gamma K_{1}\; (u) \Gamma R_{21}\; (-v,-u).
$$

Further generalizations of the YBE and the corresponding RE [37] are related to
the colored algebras \cite{Col} and/or more complicated commutation relations
among
entries of the $T$-matrices and the $K$-matrices. However the structure of
these
equations can be easily obtained following the standard ZF algebra pattern.
In particular, if the multiplicative factor $\omega (a,b)$ of
the color ZF algebra is nondegenerate \cite{Col}, where $a, b \in \cal A$
and $\cal A$ is an abelian grading group, then the graded FRT-relation is

\begin{equation}
R_{12}\; (u,v) \Gamma^{-1} T_{1}\; (u)\Gamma T_{2}\; (v) =
T_{2}\; (v) \Gamma^{-1} T_{1}\; (u)\Gamma R_{12}\; (u,v)
\end{equation}

\noindent
where $(\Gamma)_{ab;cd} = \omega (a,b) \delta_{ac} \delta_{bd}$ and we
identify for simplicity the matrix indices with the grading group elements
(which is not always the case as e.g. the $Z_2$-grading). Now $\Gamma$ is not
necessary unipotent. The corresponding RE follows from the covariance arguments

\begin{equation}
R_{12}\; (u,v) \Gamma^{-1} K_{1} \; (u)\Gamma R_{21}\; (v,- u) K_{2}\; (v) =
\end{equation}

$$
= K_{2}\; (v)R_{12}\; (u,- v) \Gamma^{-1} K_{1}\; (u) \Gamma R_{21}\; (-v,-u).
$$

\noindent
To extract the commuting functionals of the $\cal A$-graded monodromy matrix
$T$ or the corresponding Sklyanin's matrix $\cal T$ with the entries as
the homogeneous elements of the grading
the $\cal A$-graded trace \cite{Col} has to
be used
$$
t(u)= tr_{\cal A}T(u)= \sum \omega (a,a)T_{aa} (u) \;.
$$

\noindent
This trace can be considered as a particular example
of the quantum trace \cite{63} with $\Gamma$ as the $R$-matrix.
It is also possible to extend these constructions further by using
instead of the multiplicative factor $\omega (a,b)$
and/or $\Gamma$ an appropriate $R$-matrix,
however the main problem for the moment is to find an interesting example
the solution of which requires the mentioned above equations.

%\newpage

\section{Integrable models with finite degrees of freedom.}

The intensive development of the YBE and the quantum group theory was strongly
influenced by the conformal field theory.  The latter one as most of the
field theoretical integrable models has the two dimensional space-time.
However,
the rich structure of the YB-algebra and RE-algebra or
the quadratic $R$-matrix algebras with
the spectral parameter dependence permits to include into this formalism
variety of known integrable models with finite degrees of freedom
(e.g. [38-40, 41, 46]) and to find
new ones,  which are physically interesting systems in  the space of the
three (and more) dimensions.

Let us consider as an example the Kowalewski-Chaplygin-Goryachev top
(KCG top). The Hamiltonian of this model is [46].

\begin{equation}\label{5.1}
H= \frac{1}{2} (J_{1}^{2} + J^{2}_{2} + 2 J^{2}_{3}) + c_{1} x_{1} + c_{2}
x_{2} + c_{3} (x^{2}_{1} - x_{2}^{2}) + c_{4} x_{1} x_{2} + c_{5}/ x^{2}_{3} ,
\end{equation}

\noindent
where $c_{i} \,, \, i = 1, \cdots , 5$ are arbitrary constants and
$J_i$, $x_i$ are the angular momenta and coordinates.
This system is integrable provided the constraint:

\begin{equation}\label{5.2}
l= \sum^{3}_{j=1} x_{j} J_{j} =0.
\end{equation}

One gets the famous Kowalewski's top for $c_3=c_4=c_5=0$. The corresponding
auxiliary linear problem (the $L$-operator) is related to the simplest
$R$-matrix:  the Yang solution (9) for $sl(2)$ with
$c= \kappa$. The $L$-operator has quadratic dependence on the
spectral parameter $u$

\begin{equation}\label{5.3}
L(u)=
\left(
\begin{array}{ll}
y_{0} u^{2} +  y_{2} u + y_{1} & y_{4}^{+} u + y_{6}^{+}\\
y_{4}^{-} u +  y_{6}^{-} & y_{3}
\end{array}
\right)= \left( \begin{array}{cc}
                  a & b \\
                  c & d
                 \end{array} \right)(u)
\end{equation}

\noindent
where $y_{4}^{\pm} = y_{4} \pm y_{5}$ and the same for $y_{6}^{\pm}$;
$ y_{0}, ... , y_{7}$ are eight dynamical variables of the model.

The corresponding YB-algebra, generated by four entries $a(u),...,d(u)$
of the $L$-operator, has its centre generated by the $q$-determinant of the
$L$-operator (\ref{5.3}) [5, 3]:

\begin{equation}
\begin{array}{ll}
det_{q} L(u) =& a (u + i \kappa /2) d (u - i \kappa /2) - b (u + i \kappa /2)
c (u - i \kappa /2)\\
\, & = Q_{1} u^{2} + Q_{2} u - Q_{3} + \kappa^{2}/4 Q_{1} \quad ,
\end{array}
\end{equation}

\noindent
where $Q_{0} = y_{0}$, $Q_{1} = y_{0} y_{3} - y_{4}^{2} - y_{5}^{2}$,
$Q_2= y_2y_3-2y_4y_6-2y_5y_7$, $Q_3=y_6^2+y_7^2- \frac{1}{2}
 \{ y_1,y_2 \} + \frac{1}{2} \kappa^2 y_0y_3$, $\{ y_i,y_j \}=y_iy_j+y_jy_i$.

The YB-algebra for the entries of $T(u) \equiv  L(u)$ results in quadratic
commutation
relations for the dynamical variables $y_k$. The nontrivial problem
is to realize the latter ones in terms of physically significant variables.
One of the realization is given [7] by the momenta $p_i$ and coordinates
$q_i$, $i=1,2$, of the two site Toda lattice, so that the $L$-operator
(\ref{5.3}) is the product of two elementary ones $(i = 1,\,2)$

\begin{equation}
L_i(u)= \left(
\begin{array}{ll}
u-p_i & -e^{q_i} \\
e^{-q_i} & 0
\end{array} \right) \quad.
\end{equation}

The realization we are looking for defined by the generators
$J_k$, $x_k$, $k=1,2,3$ of the Lie algebra e(3), provided that
$l= \sum_{k=1}^3 x_kJ_k=0$  [46]
\begin{equation}
\begin{array}{ll}
y_0=1 & y_4= ibx_1 \\
y_1=-(J_1^2 + J_2^2 + \frac{1}{4} + 2 \alpha /x_3^2) &
y_5=ibx_2 \\
y_2= -2J_3 & y_6= - \frac{1}{2}ib \{ x_3, J_1 \} \\
y_3=b^2x_3^2 & y_7=-\frac{1}{2}ib \{ x_3, J_2 \} \quad ,
\end{array}
\end{equation}

\noindent
where $\alpha$ and $b$ are constants related to the YB-algebra central
elements.
The integrals of motion are generated by the trace of (\ref{5.3})
\begin{equation}
t(u)= tr L(u)= a(u) + d(u)= u^2-2J_3u-2H- \frac{1}{4} \quad,
\end{equation}

\noindent
where $H=\frac{1}{2}(J_1^2 + J_2^2 -b^2x_3^2) +  \alpha /x_3^2$ is the
Hamiltonian of the Newmann's systems. The solution of this system by the
separation of variables  (from $J_k$, $x_k$ to the new ones) in the
framework of the QISM is achieved using the {\bf functional Bethe Ansatz}
[7, 41]: the introduction of the new variables  as the operator roots
$\hat{u}_i$, $i=1,2$, of the entry $c(u)$ (\ref{5.3})  and the conjugated
variables
$\hat{m}_i$ as the values of $a(u)$ and $d(u)$ at these roots. Then, the
eigenvalue equation for the transfer matrix $t(u)$ (two degrees of freedom
for the Newmann's case) is reduced to the one-dimensional problems.
To embed the KCG top into such approach one has to use the RE and the
Sklyanin monodromy matrix.

The $c$-number solutions to the RE with the $sl(2)$ $R$-matrix have the form
\begin{equation}
K_{a} (u) = \alpha_{a} \sigma_{0} + u (\beta_{a}^{-} \sigma_{+} +
\beta_{a}^{+} \sigma_{-}),
\end{equation}

\noindent
where $a = \pm, \quad (\beta^{-}_{-}, \beta_{-}^{+}, \beta_{+}^{-},
\beta_{+}^{+}) = (1, - \beta_{1}, \beta_{2}, -1)$. Using the L-operator
(\ref{5.3}) one can construct according to the general recipe [29] the
monodromy matrix

\begin{equation}
{\cal T} (u) = L (u) K_{-} (u - i x/2) \sigma_2 L^T (-u)
\sigma_2
\end{equation}

\noindent
and the correspondent generating function $\tau (u)$ of the integrals
of motion

\begin{equation}
\tau (u) = tr K_{+} (u + i x/2) {\cal T} (u) \quad .
\end{equation}

\noindent
The latter one gives rise to the KCG top Hamiltonian (\ref{5.1}).

Another wide class of models embedded recently into this scheme [49]
includes the so called quasi-solvable models; some of them are
physically relevant.

{\bf Acknowledgements:} This work is supported in part by the DGICYT
(Spain). The author wishes to thank the hospitality
of the Department of Theoretical Physics of Valencia University,
H. Grosse and L. Pittner for the kind invitation to the Schladming
International Winter School of Theoretical Physics.

\end{document}